# The Deposition of High-Quality HfO$_2$ on Graphene and the Effect of Remote Oxide Phonon Scattering

Ke Zou[1], Xia Hong[1], Derek Keefer[2, 3] and Jun Zhu[1*]

[1]Department of Physics, The Pennsylvania State University, University Park, PA 16802-6300

[2]Department of Chemistry, Beloit College, Beloit, WI 53511

[3]Present Address: Department of Chemistry, The Pennsylvania State University, University Park, PA 16802-6300

*: To whom correspondence and requests for materials should be addressed, email: jzhu@phys.psu.edu.

**Abstract:**

**We demonstrate the atomic layer deposition of high-quality HfO$_2$ film on graphene and report the magnitude of remote oxide phonon (ROP) scattering in dual-oxide graphene transistors. Top gates with 30 nm HfO$_2$ oxide layer exhibit excellent doping capacity of greater than $1.5 \times 10^{13}$/cm$^2$. The carrier mobility in HfO$_2$-covered graphene reaches 20,000 cm$^2$/Vs at low temperature, which is the highest among oxide-covered graphene and compares to that of pristine samples. The temperature-dependent resistivity $\rho(T)$ exhibits the effect of ROP scattering from both the SiO$_2$ substrate and the HfO$_2$ over-layer. At room temperature, surface phonon modes of the HfO$_2$ film centered at 54 meV dominate and limit the carrier mobility to ~ 20,000 cm$^2$/Vs. Our results highlight the important choice of oxide in graphene devices.**

As graphene research rapidly advances towards electronic applications, the need to incorporate high-quality oxides into field effect transistors (FETs) and the understanding of their role in electric transport become imminent. Extensive studies in Si transistors show that charge traps and remote optical phonons of gate oxides impact the electron mobility $\mu$ in silicon channels significantly [1-5]. Similar issues arise in graphene field effect transistors (GFETs), even more prominently given graphene's atomic thickness and exposure to dual gate oxides in locally-gated structures[6, 7]. A variety of materials including $HfO_2$, $Al_2O_3$, $SiO_2$ have all been used as top-gate dielectrics [8-13], but their influences on electron transport have not been well studied. In this Letter, we demonstrate the growth of high gate-oxide quality thin $HfO_2$ film on graphene and investigate the effect of remote oxide phonon (ROP) scattering in a dual oxide $HfO_2$/graphene/$SiO_2$ structure, quantitatively assessing individual contributions from the $HfO_2$ over-layer and the $SiO_2$ substrate. At low temperature, $HfO_2$-covered graphene exhibits field effect mobility $\mu_{FE}$ up to 20,000 $cm^2$/Vs, which exceeds the highest $\mu$ reported in the literature for oxide-covered graphene (8,600 $cm^2$/Vs in Ref.10). Magneto-resistance oscillations and half-integer quantum Hall states are well developed in these samples, comparable to that of pristine exfoliated graphene. At elevated temperatures, $\mu$ decreases rapidly due to ROP scattering of the $SiO_2$ substrate and the $HfO_2$ over-layer, with the low-energy modes of the $HfO_2$ over-layer (~ 54 meV) being the dominant source. This mechanism limits $\mu$ to 20,000 $cm^2$/Vs at 300 K. These studies provide quantitative and essential information to the design and performance optimization of graphene transistors.

Conventional GFETs are fabricated using exfoliated graphene as described in Ref. 15. $HfO_2$ films are patterned and deposited on graphene using atomic layer deposition (ALD) at 110 °C using two precursors: $H_2O$ and $Hf(NMe_2)_4$ (See online supporting materials for details). Figure 1(a) shows the optical micrograph of a GFET partially covered by 30 nm of $HfO_2$. An atomic force microscope (AFM) image of the area between two metal contacts is given in Fig. 1(b). In all devices, the $HfO_2$ film grows continuously across the graphene/$SiO_2$ boundary, displaying a step height typical of single-layer graphene (7 Å in Fig. 1(b)). The film appears amorphous and smooth on both sides, providing the initial evidence of high-quality gate oxide. In addition to lithographically patterned graphene, we find that HfO$_2$ can grow on pristine exfoliated single-layer graphene directly without a seeding layer. Films thicker than 10 nm are typically pinhole free and show excellent morphology with RMS roughness of 2 – 3 Å as shown Fig. 1(c). On the other hand, films grown on multi-layer (5 – 6 layers) sheets show much poorer coverage (Fig. 1(d)), in agreement with previous studies[16-18]. This observation led us to speculate that on single-layer graphene, curvatures induced by the underlying SiO$_2$ substrate facilitate the absorption and reaction of the precursors. Details of the growth on pristine single and multi-layer graphene are given in online supporting materials.

We determine the static dielectric constant $\epsilon^0$ of the HfO$_2$ film through its gating efficiency on graphene to be $\epsilon^0 = 17 \pm 0.2$, which is consistent with values reported in previous low temperature (90 – 150 °C) growth[9,19]. Top gates using 30 nm HfO$_2$ as the dielectric layer can induce more than $1.5 \times 10^{13}$/cm$^2$ carriers into graphene, which exceed the general range of the SiO$_2$ backgate (~$1 \times 10^{13}$/cm$^2$). The high dielectric constant and excellent breakdown characteristics of HfO$_2$ enable efficient and high charge accumulation in graphene transistors.

Resistivity and quantum Hall measurements are carried out on HfO$_2$–covered GFETs in a Janis He$^4$ cryostat with a 9 T magnet using standard low-frequency lock-in techniques with current excitations ranging 50 – 100 nA. GFETs partially covered by HfO$_2$ (Fig. 1(a)) are fabricated to enable quantitative evaluation of the influences of both the SiO$_2$ substrate and the HfO$_2$ over-layer. Figure 2(a) plots the low-T conductivity $\sigma(V_{bg})$ on the HfO$_2$ side of three partially covered GFETs (A – C). The low-density field effect mobility $\mu_{FE} = (d\sigma/dn)(1/e)$ is determined to be 9,600, 17,000 and 11,200 cm$^2$/Vs respectively. These values are close to $\mu_{FE}$ on the bare side of the same device, which are 11,500, 16,100 and 10,400 cm$^2$/Vs respectively[20].
We estimate $\mu_{FE}$ to be above 20,000 cm$^2$/Vs in a 2-terminal GFET (D) shown in Fig. 2 (c). These $\mu_{FE}$ values compare or exceed the best $\mu$ = 8,600 cm$^2$/Vs reported for oxide-covered graphene[10]. In addition to high $\mu$, HfO$_2$-covered graphene exhibits well-developed half-integer quantum Hall states and magneto-resistance oscillations, similar to that of pristine exfoliated graphene[23-25] (Figs. 2 (b) and (c)). These observations provide additional evidence for the high quality of the samples.

At high temperature, electrons in graphene are subject to scattering by the electric field generated by the polar optical phonon modes in a nearby oxide layer. First discussed by Wang and Mahan[1], this mechanism is known to be an important mobility limiting factor in silicon transistors, especially when high-$\kappa$ oxides with low-energy phonon modes are used[2,4]. Recent studies indicate that it may also be responsible for the rapid increase of the resistivity with temperature in graphene-on-SiO$_2$ samples at $T > 100$ K[6,14]. We measure and compare the $T$-dependent resistivity $\rho(T)$ in three partially covered GFETs (A-C) on both bare and HfO$_2$-covered graphene in the range of 10 K $< T <$ 250 K[26] and for carrier densities $1 \times 10^{12}$/cm$^2 < n < 3 \times 10^{12}$/cm$^2$. Figures 3(a) (bare) and (b) (covered) plot the $\rho(T)$ data of sample B separately and Figure 3(c) plots $\rho(T)$ on both sides at $n = 3 \times 10^{12}$/cm$^2$ for comparison. For 10 K $< T <$ 100 K, $\rho(T)$ on both sides increases linearly with T with the same slope. Previously this linear-T dependence was attributed to longitudinal acoustic (LA) phonon scattering

$\rho_{LA}(T) = (h/e^2)\pi^2 D_A^2 k_B T / (2h^2 \rho_S v_S^2 v_F^2)$ [6,27]. Our data support this conclusion. Fitting to Eq. (1) described below yields a deformation potential $D_A = 18 \pm 2$ eV for all samples, in good agreement with $D_A = 18 \pm 1$ eV obtained by Chen et al.[6].

Above 100 K, $\rho(T)$ increases more rapidly with temperature, with a higher rate on the HfO$_2$-covered side. The interpretation of this rapid rise is still controversial. While some experiments[6] and calculations[14] pointed to remote substrate phonon scattering $\rho(T)$, others have proposed the thermal activation of quenched ripples[7]. Here we examine this issue independently and investigate possible contributions from additional phonon modes of the HfO$_2$ over-layer. We compare our data to the model of remote oxide phonon scattering only, since a quantitative theory of quenched ripples does not exist. $\rho(T, n)$ is analyzed using the following equation:

$$\rho(T, n) = \rho_0(n) + \rho_{LA}(T) + \rho_{ROP}(T, n), \quad (1)$$

where $\rho_0(n)$ represents the low-$T$ residual resistivity, $\rho_{LA}(T)$ is the LA phonon contribution described earlier and $\rho_{ROP}(T, n)$ originates from remote oxide phonon scattering. $\rho_{ROP}(T, n)$ is given by:

$$\rho_{ROP}(T,n) = \int A(\vec{k},\vec{q}) d\vec{k} d\vec{q} \sum_i g_i / \left(e^{\hbar\omega_i/k_B T} - 1\right), \quad (2)$$

where $A(\vec{k},\vec{q})$ is the matrix element of scattering events between electron ($\vec{k}$) and phonon ($\vec{q}$) states and $\omega_i$ and $g_i$ represent the frequency and strength of the i$^{th}$ ROP mode respectively[1,4,14].

On the vacuum/graphene/SiO$_2$ side of the GFET, two ROP modes from the SiO$_2$ substrate are found to be important[6,14]. Using equations given in Fischetti et al.[4] and dielectric constants measured for our substrates, we determine the frequency and strength of the surface modes and obtain $\omega_1$ = 63 meV, $\omega_2$ = 149 meV, $g_1$ = 3.2 meV and $g_2$ = 8.7 meV respectively. Details of the calculation are given in the online supporting materials.

Fittings to Eq. (1) provide excellent description of $\rho(T, n)$ on the bare side of the GFET (Fig. 3(a)). Contributions from the $\omega_2$ mode are negligible in the temperature range studied, as expected from Eq. (2) and verified by the fittings. The resulting $T$-independent resistivity coefficient of the $\omega_1$ mode $C_1(n) = g_1 \int A(\vec{k},\vec{q}) d\vec{k} d\vec{q}$ is plotted in Fig. 4(a) for samples A – C. At

300 K, the $\omega_1$ mode is projected to produce a resistance of ~30 Ω at $n = 3 \times 10^{12}$/cm$^2$ and follows an approximate $1/n$ density dependence. $C_1(n)$ varies less than 25 % among our samples, including conventional graphene-on-SiO$_2$ devices not shown here. They are also in excellent agreement with values reported by Chen et al.[6] The consistency and reproducibility of the experimental observations are quite remarkable. Resistances calculated in Ref. 14 are roughly 50 % of experiments. These agreements are supportive of the ROP scattering model.

$\rho(T, n)$ on the HfO$_2$-covered side of a GFET consistently exhibits stronger $T$-dependence (Fig. 3(b)), suggesting the presence of another scattering channel. This observation is difficult to reconcile within the quenched ripple scenario. Instead, we consider possible surface phonon modes introduced by the HfO$_2$ over-layer. Indeed, the IR adsorption spectra of our HfO$_2$ films exhibit a broad maximum centered at 40 meV (320 cm$^{-1}$) (Fig. 3(d)), due to its amorphous nature[28], whereas crystalline HfO$_2$ possesses well-defined TO and LO modes in the 100-700 cm$^{-1}$ range[4]. We approximate the corresponding distribution of surface phonon modes with a single frequency $\omega_3$, where $\omega_3$ = 54 meV is calculated from the average bulk TO mode frequency $\omega_{TO}$ = 40 meV, as given by the IR data. In addition to the new surface mode, the presence of the HfO$_2$ over-layer also modifies the frequency of the existing SiO$_2$ surface modes slightly due to the new boundary condition $\epsilon_{SiO2}(\omega) + \epsilon_{HfO2}(\omega) = 0$ and significantly reduces the strength of the modes $g_1$, $g_2$ due to electronic screening of the HfO$_2$ over-layer. On the HfO$_2$-covered side, we obtain $\omega_1'$ = 72 meV, $g_1'$ = 1.2 meV (SiO$_2$), $\omega_2'$ = 143 meV, $g_2'$ = 2.4 meV (SiO$_2$) and $\omega_3$ = 54 meV, $g_3$ = 5.7 meV (HfO$_2$). Details of the calculation are given in the online supporting materials.

$\rho(T, n)$ on the covered side of the GFETs (A-C) are fit to Eq. (1) considering two surface phonon modes $\omega_1'$ and $\omega_3$. The two frequencies are too close to be differentiated by the fitting itself. Instead, $C_1'(n) = C_1(n)/2.6$ is used as input to Eq. (1) to extract $C_3(n)$. The fittings describe data very well as shown in Fig. 3(b). The resulting $C_3(n)$ is plotted in Fig. 4(b). Despite the crude approximation used to describe the phonon modes in $HfO_2$, the extracted mode strength ratio $C_3(n)/C_1'(n)$ ranges 2.5 – 5.5 and reaches good agreement with the calculated ratio $g_3/g_1' = 4.7$.

The above analyses attest to the success of the ROP scattering model in explaining the magnitude and $T$-dependence of $\rho(T, n)$ in $HfO_2$/graphene/$SiO_2$ structures. Questions remain, however, on the $n$-dependence of $C_1(n)$ and $C_3(n)$, as well as the variation of $C_3(n)$ among samples. These issues are briefly addressed in the online supporting materials, where we discuss the treatment of electron screening and the effect of a spacing layer at the graphene-oxide interface. To fully understand the role of oxide in GFETs, the understanding and control of the graphene-oxide interface proves essential.

In Fig. 5, we summarize the magnitude of the various phonon scattering mechanisms by plotting $\mu_i(n) = 1/(ne\rho_i(n))$ corresponding to the resistance due to each phonon channel in $HfO_2$/graphene/$SiO_2$ GFETs. At 300 K, the LA phonon of graphene contributes to a constant resistivity of 30 $\Omega$, resulting in an $n$-dependent $\mu$ that is approximately $1 \times 10^5$ cm$^2$/Vs at $n = 2 \times 10^{12}$/cm$^2$. The ROP modes of the $SiO_2$ substrate, while limiting $\mu$ to ~ 60,000 cm$^2$/Vs in back-gate only devices, play a minor role in $HfO_2$/graphene/$SiO_2$ devices due to the screening of the $HfO_2$ over-layer, giving rise to $\mu \sim 2 \times 10^5$ cm$^2$/Vs. At 300 K, the most dominant phonon scattering source comes from the ROP mode of the $HfO_2$ over-layer, which limits $\mu$ to approximately 20,000 cm$^2$/Vs. These results provide key insight towards the design of graphene electronics. While high-$\kappa$ oxides such as $HfO_2$ enable efficient carrier injections, their negative influence on carrier mobility must be taken into account.

In conclusion, we demonstrate the atomic layer deposition of high-quality $HfO_2$ film on graphene and report the highest mobility $\mu \sim$ 10,000 – 20,000 cm$^2$/Vs among oxide-covered GFETs.

Remote surface phonons of the $HfO_2$ film lead to strong scattering at high temperature and limit carrier mobility in graphene to 20,000 $cm^2$/Vs at 300 K. Our results highlight the importance of oxide choice in GFETs. The methods and analyses employed here may be generalized to examine the effect of other substrates and top-gate oxides in single- and double-gated FETs.

**Acknowledgements:**

We are grateful for helpful discussions with Vin Crespi, Peter Eklund, Simone Fratini, Jainendra Jain and Jerry Mahan. The IR adsorption spectra are taken with the assistance of Xiaoming Liu. This work is supported by NSF CAREER DMR-0748604, NSF NIRT ECS-0609243. D. K. acknowledges the support of the NSF NNIN REU-0335765. The authors acknowledge use of facilities at the PSU site of NSF NNIN.

**Figure 1:** The growth of $HfO_2$ on pristine graphene and partially covered field effect transistors. (a) Optical micrograph of a partially covered GFET. $HfO_2$ film appears green in the image. Film thickness is 30 nm. The graphene sheet is outlined in red. (b) AFM image of the circled area in (a) showing the continuous growth of $HfO_2$ over the graphene (left) / $SiO_2$ (right) step. A line cut across the step indicates a height of 7 Å. The surface RMS roughness is ~ 3 – 4 Å on graphene and ~ 2 – 3 Å on $SiO_2$. (c) and (d) AFM image of 10 nm $HfO_2$ grown on pristine single-layer (c) and multi-layer (5 – 6 layers) graphene (d). The RMS roughness of the film is ~ 2 – 3 Å on single layer graphene. See the online Supplementary Information for more details.

**Figure 2:** Transport properties of $HfO_2$-covered graphene. (a) Low-$T$ conductivity $\sigma(V_{bg})$ of $HfO_2$-covered graphene in three partially covered GFETs. The low-density field effect mobility $\mu_{FE}$ is determined to be 9,600 cm$^2$/Vs (Sample A, black), 17,000 cm$^2$/Vs (Sample B, red) and 11,200 cm$^2$/Vs (Sample C, blue) for the carrier type with higher mobility. $\mu_{FE}$ on the bare side of the same devices reaches 11,500, 16,100 and 10,400 cm$^2$/Vs respectively (not shown). The majority of our $HfO_2$-covered samples exhibit Dirac points within ± 20 V and $\mu_{FE}$ > 6000 cm$^2$/Vs. Some samples show electron-hole asymmetry, likely due to contacts, as in pristine graphene samples[29]. (b) Well developed half-integer quantum Hall states in sample A at $B$ = 8.9 T and $T$ = 1.5 K. (c) Shubnikov-de Hass oscillations in a fully covered 2-terminal GFET (sample D) at $n$ = 3.6x10$^{12}$/cm$^2$ (background subtracted). We estimate $\mu_{FE}$ to be ~ 20,000 cm$^2$/Vs in this device. The quantum scattering time derived from the oscillations $\tau_q$ = 85 fs is comparable to the best pristine graphene samples[25]. Inset of (d): Raw $R_{xx}$ (blue solid) and the background (red dash).

**Figure 3:** Remote oxide phonon (ROP) scattering in bare and HfO$_2$-covered graphene. (a)(b) Resistivity $\rho(T)$ of bare (a) and HfO$_2$-covered (b) graphene in sample B. From top to bottom: hole density $n = 1.0 - 3.0 \times 10^{12}$/cm$^2$ in $5.0 \times 10^{11}$/cm$^2$ steps. Error bar is smaller than symbol size. (c) $\rho(T)$ of sample B on the bare (black) and covered (red) side shown together for comparison. $n = 3.0 \times 10^{12}$/cm$^2$. Both sides show the same linear T-dependence below 100 K, which has a slope of 0.1 Ω/K and leads to a LA phonon deformation potential of $D_A = 18$ eV. Fittings at other densities use the same slope. $\rho(T)$ of the HfO$_2$-covered side increases more rapidly at higher temperature. Solid lines in (a) – (c) are fittings to Eq. (1) including $\rho_0$, LA and ROP phonons. (d) Top: Normalized FTIR adsorption spectra of 55 nm HfO$_2$ grown on intrinsic Si substrate. The adsorption coefficient reaches a maximum of 11 % at 322 cm$^{-1}$ or 40 meV, which is identified as the average energy of the bulk TO mode. Bottom: background signal of the intrinsic Si substrate.

**Figure 4:** The amplitude of scattering by surface optical phonon modes of SiO$_2$ and HfO$_2$. (a) Resistivity coefficient vs. density $C_1(n)$ of the $\omega_1 = 63$ meV mode of the SiO$_2$ surface on the bare side of three GFETs. $C_1'(n)$ of the same mode on the HfO$_2$-covered side decreases by a factor of 2.6 due to screening of the HfO$_2$ over-layer. The contribution of the $\omega_2 = 149$ meV mode is negligible in the temperature range studied. (b) $C_3(n)$ of the $\omega_3 = 54$ meV mode of the HfO$_2$ surface on the covered side of the same GFETs. Open squares, solid circles and solid triangles correspond to samples A-C respectively. Dashed lines are empirical fittings to $n^{-\alpha}$, where $\alpha$ range 1.1 – 1.6 for $C_1(n)$ and 0.7 – 1.0 for $C_3(n)$. Electrons and holes exhibit similar $C(n)$. Only one carrier type is shown for each sample.

**Figure 5:** Mobility limit imposed by LA and ROP scattering at 300 K. $\mu$ is limited to 50,000-70,000 cm$^2$/Vs by the $\omega_1 = 63$ meV mode in graphene-on-SiO$_2$ samples (black) and 170,000-220,000 cm$^2$/Vs in HfO$_2$/graphene/SiO$_2$ structures (magenta). The $\omega_3 = 54$ meV mode in HfO$_2$ covered GFETs limits $\mu$ to ~20,000 cm$^2$/Vs (red). The blue line represents $\mu$ set by LA phonons. Sample B, which exhibits the highest phonon scattering among all devices, is used to generate this figure.

Figure 1

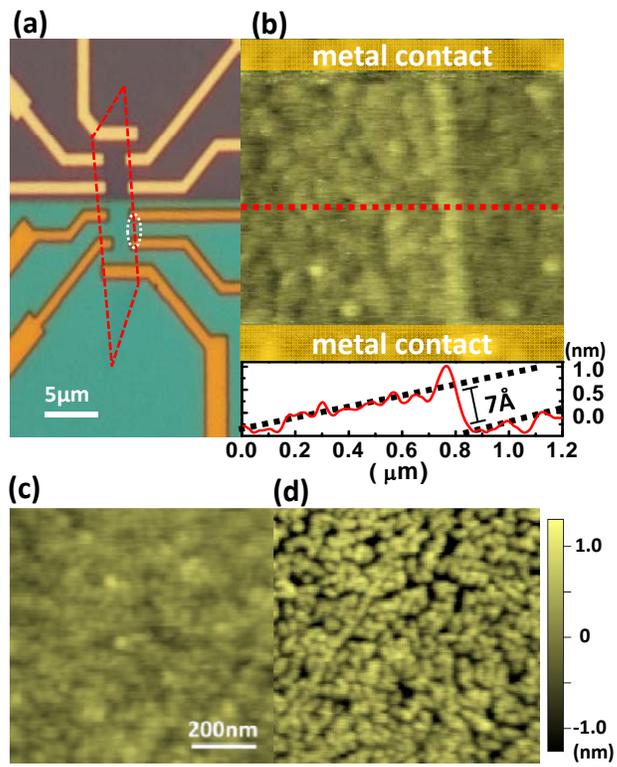

Figure 2

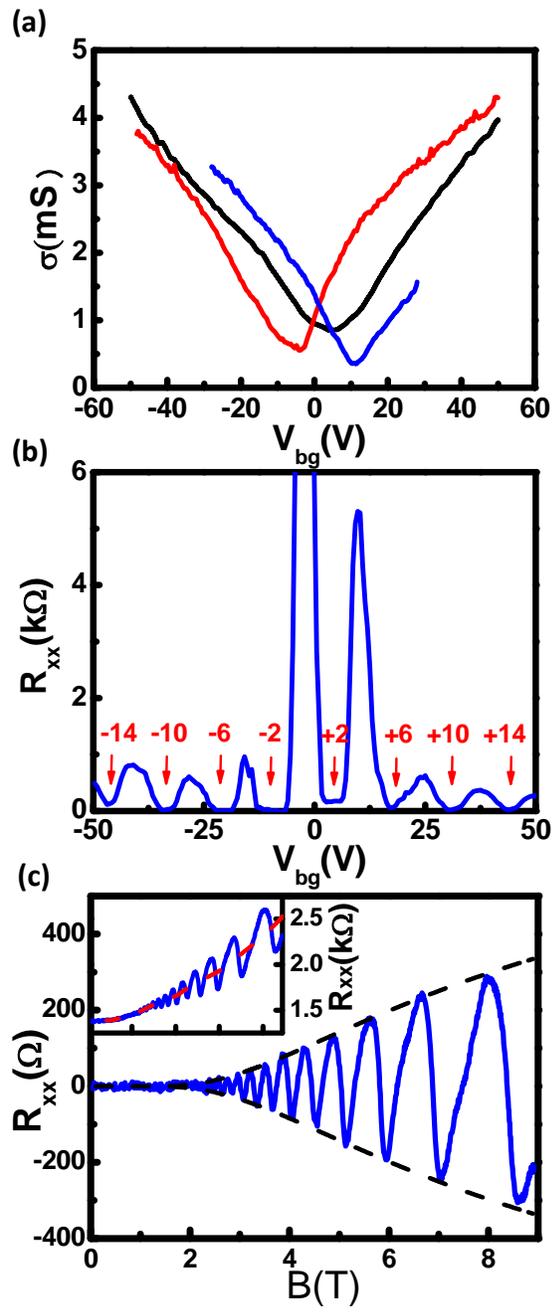

Figure 3

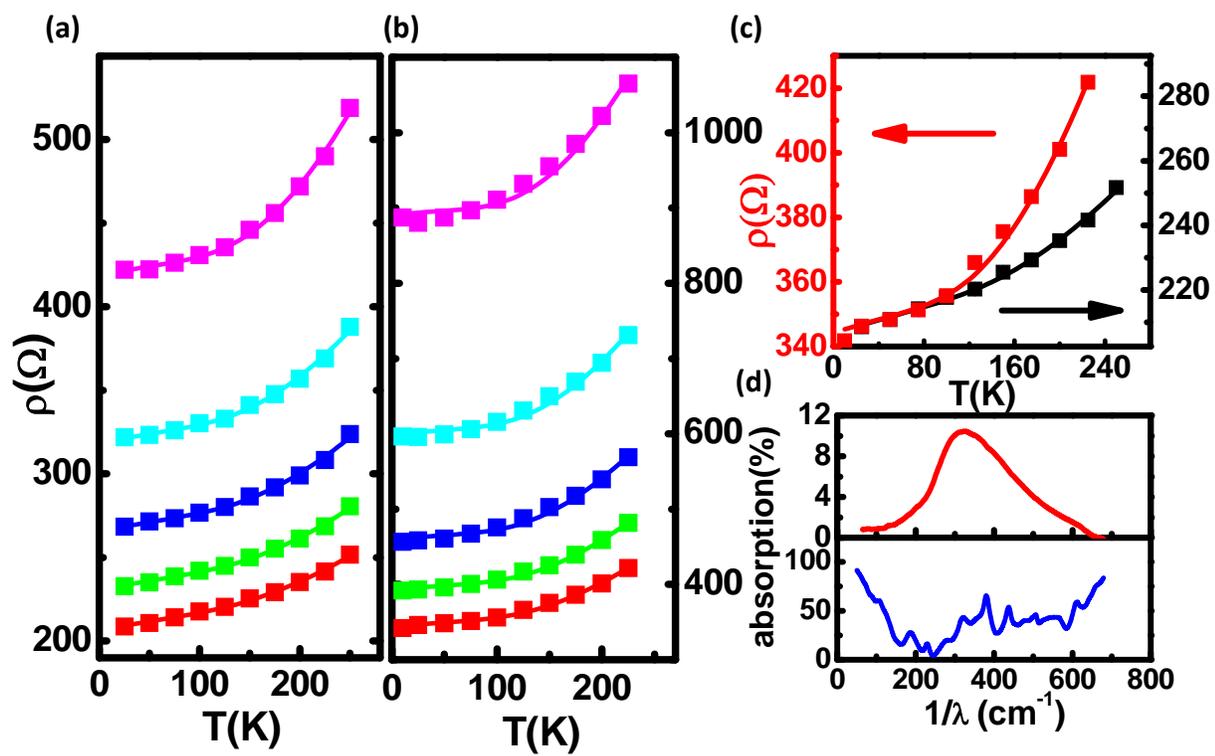

Figure 4

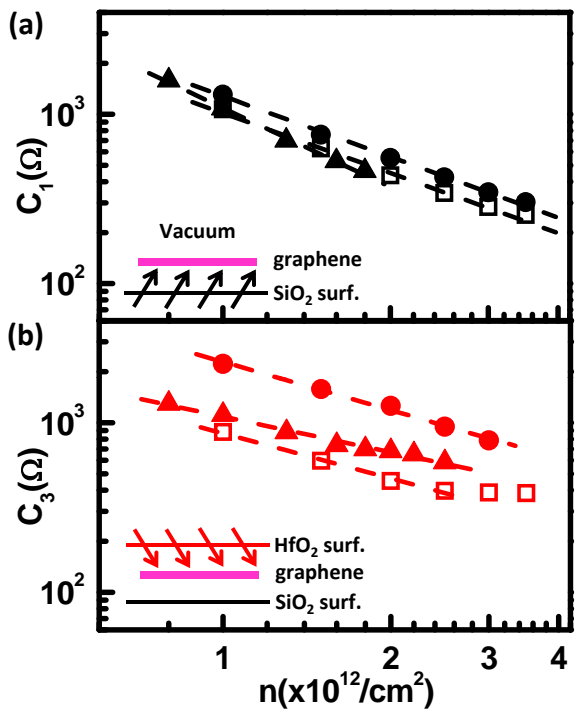

Figure 5

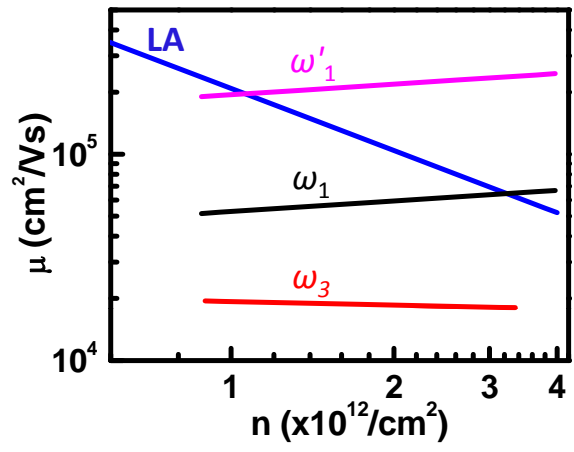

# The Deposition of High-Quality $HfO_2$ on Graphene and the Effect of Remote Oxide Phonon Scattering

Ke Zou[1], Xia Hong[1], Derek Keefer[2, 3] and Jun Zhu[1*]

[1]Department of Physics, The Pennsylvania State University, University Park, PA 16802-6300

[2]Department of Chemistry, Beloit College, Beloit, WI 53511

[3]Present Address: Department of Chemistry, The Pennsylvania State University, University Park, PA 16802-6300

*email: jzhu@phys.psu.edu.

## Supporting Materials

### 1. Device fabrication

Single layer graphene flakes are prepared by exfoliating HOPG (ZYA grade, SPI supplies) onto $SiO_2$ (290 nm)/doped Si wafer. We use standard e-beam lithography and metal deposition techniques (Ti 5 nm/ Au 50 nm) to process rectangular pieces into hall bar devices. A second e-beam writing exposes the area to be covered by $HfO_2$. Immediately following the development of the resist, $HfO_2$ films are deposited in a Cambridge Savannah 200 Atomic Layer Deposition system using two precursors: $H_2O$ and $Hf(NMe_2)_4$ (Sigma-Aldrich). The chamber temperature is 110 °C and the growth rate is ~ 1.2 Å/cycle. The device is then soaked in warm acetone (~ 40 °C) for 15 – 30 mins, with the aid of ultrasound sonification up to 5 mins when necessary, to release the unwanted $HfO_2$ film. On some devices, a third e-beam lithography and metal deposition are used to fabricate the top-gate electrode.

### 2. Growth of $HfO_2$ on pristine graphene

Previously, while some groups have successfully deposited continuous oxide layers on exfoliated graphene using ALD without the need of an adhesion layer[1], others have reported the necessity of such a layer (NCFL, HSQ, Al, PTCA or $O_3$)[2-7] to achieve smooth $HfO_2$ or $Al_2O_3$ films. Indeed, the growth of $HfO_2$ on pristine graphene appears puzzling owing to its lack of dangling bonds. To identify the growth mechanism, we have deposited and imaged $HfO_2$ films of varying thickness on pristine single- and multi-layer (5-6) graphene flakes exfoliated on $SiO_2$ substrate.

On single-layer graphene, $HfO_2$ films with $d$ = 2.5 nm show a coverage of ~98 % with visible pinholes. Films thicker than 10 nm are pinhole free and show excellent surface morphology with RMS roughness of 2-3 Å, comparable to that of the $SiO_2$ substrate. $HfO_2$ films deposited on thicker graphene flakes concurrently exhibit poorer quality consistently. The coverage of 2.5 nm films on 5-6 layer graphene is only about 50 % and pinholes remain in 20 nm films. These observations, together with the fact that the surface of single-layer graphene deposited on $SiO_2$ is significantly rougher than those of multi-layers, led us to speculate that the existing curvature in single-layers facilitates the adsorption and reaction of the precursors. This hypothesis can potentially explain the low coverage also observed in Refs. 5-7, where the growth was done on ~ 5-layer graphene or HOPG. We note that in graphene field effect transistor (GFET) devices a thin layer of e-beam resist residue (~ few Å) after the developing process may exist at the graphene-oxide interface so that the growth mechanism there may be different (see discussions in Section 4).

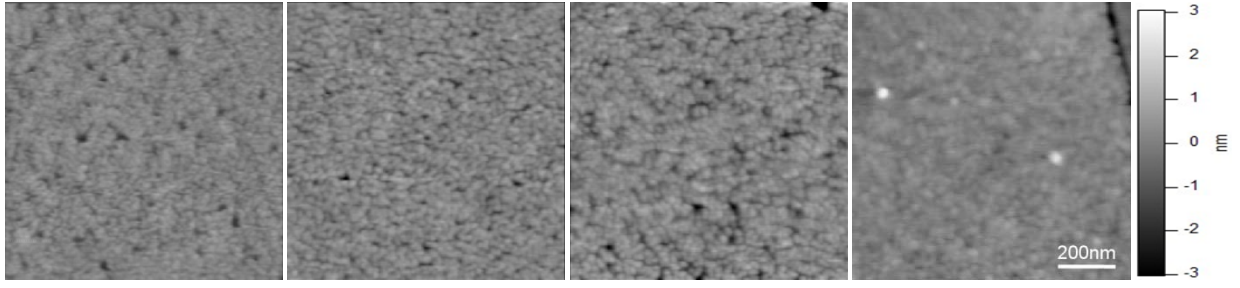

**Supplementary Figure 1:** The growth of HfO$_2$ on exfoliated single layer graphene at 110 °C. (left to right) AFM images of HfO$_2$ films with thickness $d$ = 2.5 nm, 5.0 nm, 7.5 nm, and 10.0 nm.

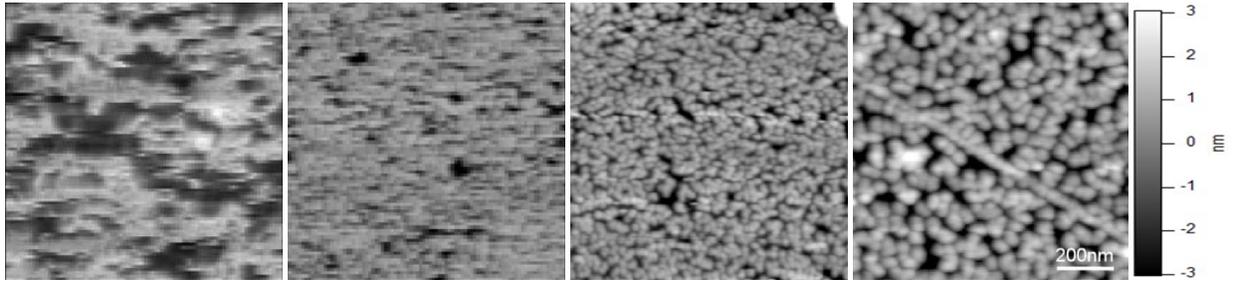

**Supplementary Figure 2:** The growth of HfO$_2$ on exfoliated multi-layer graphene at 110 °C. (left to right) AFM images of HfO$_2$ films with thickness $d$ = 2.5 nm, 5.0 nm, 7.5 nm, and 20.0 nm.

## 3. Frequencies of the surface phonon modes

Following Fischetti et al. (Ref. 8), the frequency-dependent dielectric constant of SiO$_2$ can be written as follows :

$$\epsilon_{SiO2}(\omega) = \epsilon_{SiO2}^{\infty} + \left(\epsilon_{SiO2}^{int} - \epsilon_{SiO2}^{\infty}\right)\frac{\omega_{TO2}^2}{\omega_{TO2}^2 - \omega^2} + \left(\epsilon_{SiO2}^{0} - \epsilon_{SiO2}^{int}\right)\frac{\omega_{TO1}^2}{\omega_{TO1}^2 - \omega^2}, \qquad (1)$$

where $\epsilon_{SiO2}^{\infty}$, $\epsilon_{SiO2}^{int}$, and $\epsilon_{SiO2}^{0}$ are the optical, intermediate and static dielectric constants of $SiO_2$ respectively, and $\omega_{TO1}, \omega_{LO1}, \omega_{TO2}, \omega_{LO2}$ are the frequencies of the transverse and longitudinal optical (TO and LO) phonon modes of the bulk $SiO_2$.

Previous studies have demonstrated the importance of two surface optical phonon modes in electron-phonon scattering[8-10], which originate from the bulk TO and LO modes. In this work, we adopt the frequencies determined in Ref. 7 from FTIR data for amorphous $SiO_2$. These are $\omega_{TO1} = 57.2$ meV, $\omega_{LO1} = 66.0$ meV and $\omega_{TO2} = 133.4$ meV, $\omega_{LO2} = 155.7$ meV respectively. The static and optical dielectric constant of our $SiO_2$ substrates are determined to be $\epsilon_{SiO2}^{0} = 3.9$ and $\epsilon_{SiO2}^{\infty} = 2.2$ respectively from low-frequency capacitance and ellipsometry ($\lambda = 632.8$ nm) measurements. The intermediate dielectric constant $\epsilon^{int} = 2.94$ is calculated from the frequency splitting between the TO and LO modes, using the generalized Lyddane-Sachs-Teller relation[8]. Both pairs yield the same $\epsilon^{int}$. The frequencies of the surface modes are calculated using Supplementary Eq. (1) and the boundary condition $\epsilon_{SiO2}(\omega) + 1 = 0$, corresponding to the vacuum/graphene/$SiO_2$ side of the GFETs. We obtain $\omega_1 = 63$ meV and $\omega_2 = 149$ meV.

Similarly, $\epsilon_{HfO2}(\omega)$ of the $HfO_2$ over-layer can be written as:

$$\epsilon_{HfO2}(\omega) = \epsilon_{HfO2}^{\infty} + \left(\epsilon_{HfO2}^{0} - \epsilon_{HfO2}^{\infty}\right)\frac{\omega_{TO}^2}{\omega_{TO}^2 - \omega^2}, \qquad (2)$$

where $\epsilon_{HfO2}^{0}$ and $\epsilon_{HfO2}^{\infty}$ are the optical and static dielectric constants of $HfO_2$ respectively, and we have approximated the bulk $HfO_2$ phonons with a single mode of $\omega_{TO} = 40.0$ meV obtained in FTIR measurements (see Fig. 3 (d) in the text). The static and optical dielectric constant of $HfO_2$ films are determined to be $\epsilon_{HfO2}^{0} = 17$ and $\epsilon_{HfO2}^{\infty} = 4.1$ respectively. On the $HfO_2$/graphene/$SiO_2$ side of the GFET, the frequencies of the surface modes are determined by the boundary condition $\epsilon_{SiO2}(\omega) + \epsilon_{HfO2}(\omega) = 0$. We obtain $\omega_1' = 72$ meV (SiO$_2$), $\omega_2' = 143$ meV (SiO$_2$) and $\omega_3 = 54$ meV (HfO$_2$).

## 4. The strength of the surface modes $C(n)$

The resistance due to remote oxide phonon (ROP) scattering is given by

$$\rho_{ROP}(T,n) = \sum_i C_i(n)/(e^{\hbar\omega i/k_B T} - 1), \tag{3}$$

where

$$C_i(n) = g_i \int A(\vec{k}, \vec{q}) d\vec{k} d\vec{q}, \tag{4}$$

and $g_i$ represents the strength of the i$^{th}$ surface mode (see text).

On the vacuum/graphene/SiO$_2$ side of the GFET, the strength of the $\omega_1$ and $\omega_2$ modes is approximately given by the static, intermediate and optical dielectric constants of SiO$_2$[8,10].

$$g_1 = \hbar\omega_1[1/(\epsilon_{SiO2}^{int} + 1) - 1/(\epsilon_{SiO2}^0 + 1)], \tag{5}$$

$$g_2 = \hbar\omega_2[1/(\epsilon_{SiO2}^{\infty} + 1) - 1/(\epsilon_{SiO2}^{int} + 1)]. \tag{6}$$

We obtain $g_1 = 3.2$ meV, $g_2 = 8.7$ meV and $C_1(n)/C_2(n) = g_1/g_2 = 1/2.8$.

On the HfO$_2$/graphene/SiO$_2$ side of the GFET, the presence of the HfO$_2$ over-layer significantly reduces the electric field of these modes due to the electronic screening. This effect is accounted for by replacing $\epsilon_{vacuum}^{\infty} = 1$ with $\epsilon_{HfO2}^{\infty} = 4.1$ [8,10]. The strength of the resulting surface modes thus reads:

$$g'_1 = \hbar\omega'_1[1/(\epsilon_{SiO2}^{int} + \epsilon_{HfO2}^{\infty}) - 1/(\epsilon_{SiO2}^{0} + \epsilon_{HfO2}^{\infty})], \quad (7)$$

$$g'_2 = \hbar\omega'_2[1/(\epsilon_{SiO2}^{\infty} + \epsilon_{HfO2}^{\infty}) - 1/(\epsilon_{SiO2}^{int} + \epsilon_{HfO2}^{\infty})], \quad (8)$$

$$g_3 = \hbar\omega_3[1/(\epsilon_{HfO2}^{\infty} + \epsilon_{SiO2}^{\infty}) - 1/(\epsilon_{HfO2}^{0} + \epsilon_{SiO2}^{\infty})]. \quad (9)$$

We obtain $g'_1$ = 1.2 meV (SiO$_2$), $g'_2$ = 2.4 meV (SiO$_2$) and $g_3$ = 5.7 meV (HfO$_2$). The strength of the $\omega'_1$ mode of SiO$_2$ is reduced by nearly three-fold, $C'_1(n)/C_1(n) = g'_1/g_1 = 1/2.6$.

A more accurate calculation of $g_i$ following Eqs. 31-34 in Ref .8 yields $g_1$ = 3.0 meV, $g_2$ = 9.5 meV, $g'_1$ = 1.3 meV and $g'_2$ = 2.5 meV, in excellent agreement with the approximate expressions given in Supplementary Eqs. 5-8 above.

## 5. The density and sample dependence of $C(n)$

In this section, we discuss the origin of the density dependence observed in $C_1(n)$ and $C_3(n)$ and the variation of $C_3(n)$ among samples. We fit the density dependence of $C_1(n)$ and $C_3(n)$ with $n^{-\alpha}$ empirically. The exponent $\alpha$ ranges $1.1 - 1.6$ for $C_1(n)$ and $0.7 - 1.0$ for $C_3(n)$ (see Fig. 4 in text). These exponents are consistent with $\alpha = 1.04$ of the $\omega_1$ mode determined by Chen et al.[9]. The value of $\alpha$ is sensitive to small variations of $C(n)$ since the density range is small. Nonetheless, these values are higher than $\alpha = 1/2$ predicted by calculations using Thomas-Fermi approximation for electron screening[10]. Thomas-Fermi approximation may overestimate the effect of screening, particularly at low densities. Therefore, a more sophisticated treatment may lead to a stronger density dependence, reducing the discrepancy between theory and experiment.

In Fig. 4(b), $C_3(n)$ displays a factor of three variation among the samples studied. One possible explanation is the existence of a spacing layer at the graphene-oxide interface . A spacing layer of thickness $d$ leads to an exponential decay of the scattering amplitude $e^{-2qd}$ [10], where $q$ is the phonon momentum. A simple estimate shows that for a characteristic $q = k_F$ at $n = 3 \times 10^{12}/cm^2$, a spacing layer of $d = 1.7$ nm can lead to a factor of 3 decay. This q-dependent exponential decay also produces an apparent density dependence $n^{-0.45}$ in the range $1 \times 10^{12}/cm^2 < n < 3 \times 10^{12}/cm^2$. In a control experiment, AFM measurements showed a 4 Å resist layer on graphene after the e-beam exposure and developing of the $HfO_2$ pattern. This observation supports the spacing layer scenario. However, a 4 Å spacing layer can only produce an attenuation of 20% and a residue layer of 1.7 nm is unlikely. In addition to the difficulty of explaining the magnitude of the variations, the spacing layer mechanism also implies that samples with thicker spacing layers should exhibit $C(n)$ of smaller magnitude and larger exponent $\alpha$ simultaneously, whereas in our data $C_3(n)$ of sample A is roughly 1/3 of that of sample B, but has nearly identical exponents. This observation suggests that a spacing layer may not be the sole reason behind the observed variations of $C_3(n)$. Indeed, despite the use of nominally identical recipes, we cannot rule out intrinsic variations among the $HfO_2$ films deposited in different runs, which may also affect the strength of the surface modes.